\documentclass[journal]{IEEEtran}
\usepackage{blindtext}
\usepackage[utf8]{inputenc}
\usepackage[english]{babel}
\usepackage{multirow}
\usepackage{amssymb}
\usepackage[noend]{algorithmic}
\usepackage{algorithm}
\usepackage{cite}
\usepackage{array}
\usepackage{textcomp}
\usepackage{stfloats}
\usepackage[final]{graphicx}
\usepackage{xcolor}
\usepackage{amsmath}
\usepackage{mathtools}
\usepackage{comment}
\usepackage{color,soul}

\DeclareMathOperator{\EX}{\mathbb{E}}% expected value
\usepackage{stackengine}

\usepackage{booktabs}
\usepackage{caption}
\usepackage[cmintegrals]{newtxmath}
\DeclareMathOperator*{\minimize}{minimize}
\usepackage{balance}
\usepackage[export]{adjustbox}
\usepackage[caption=false,font=normalsize=sf,textfont=sf]{subfig}
\usepackage{lipsum} %% gerador
\usepackage{siunitx}
\makeatletter
\providecommand\add@text{}
\newcommand\tagaddtext[1]{%
  \gdef\add@text{#1\gdef\add@text{}}}% 
\renewcommand\tagform@[1]{%
  \maketag@@@{\llap{\add@text\quad}(\ignorespaces#1\unskip\@@italiccorr)}%
}
\makeatother
\newcommand{\mybibliography}{\bibliography{jour_full,conf_full, references.bib}}

\begin{document}
\bstctlcite{IEEEexample:BSTcontrol}

\title{Federated Learning-Distillation Alternation for Resource-Constrained IoT}%

\author{Rafael~Valente~da~Silva,
Onel~L.~Alcaraz~López,
and~Richard~Demo~Souza%% <-this % stops a space
\thanks{This work was supported by 6G Flagship (Grant Number 369116) funded by the Research Council of Finland, the European Commission through the Horizon Europe/JU SNS project Hexa-X-II (Grant Agreement no. 101095759), CNPq (305021/2021-4) and RNP/MCTI Brasil 6G (01245.020548/2021-07).}% <-this % stops a space
\thanks{R. V. Silva and R. D. Souza are with the Department of Electrical and Electronics Engineering of the Federal University of Santa Catarina, Florianópolis, Brazil \{r.valente@posgrad.ufsc.br, richard.demo@ufsc.br\}.}%
\thanks{Onel L. Alcaraz López is with the Centre for Wireless Communications (CWC), University of Oulu, 90570 Oulu, Finland \{onel.alcarazlopez@oulu.fi\}.}}%

\maketitle

\begin{abstract}
Federated learning (FL) faces significant challenges in Internet of Things (IoT) networks due to device limitations in energy and communication resources, especially when considering the large size of FL models. From an energy perspective, the challenge is aggravated if devices rely on energy harvesting (EH), as energy availability can vary significantly over time, influencing the average number of participating users in each iteration. Additionally, the transmission of large model updates is more susceptible to interference from uncorrelated background traffic in shared wireless environments. As an alternative, federated distillation (FD) reduces communication overhead and energy consumption by transmitting local model outputs, which are typically much smaller than the entire model used in FL. However, this comes at the cost of reduced model accuracy. Therefore, in this paper, we propose FL-distillation alternation (FLDA). In FLDA, devices alternate between FD and FL phases, balancing model information with lower communication overhead and energy consumption per iteration. We consider a multichannel slotted-ALOHA \mbox{EH-IoT} networ ksubject to background traffic/interference. In such a scenario, FLDA demonstrates higher model accuracy than both FL and FD, and achieves faster convergence than FL. Moreover, FLDA achieves target accuracies saving up to $\boldsymbol{98}$\% in energy consumption, while also being less sensitive to interference, both relative to FL.
\end{abstract}
\section{Introduction}
\label{sec:intro}
\IEEEPARstart{T}{he} proliferation of the Internet of Things (IoT) paradigm as part of advancements towards $6$G is intensifying data collection and processing. Notably, federated learning (FL) has become an attractive option in a variety of IoT applications, ranging from smart healthcare \cite{Myrzashova2023} to unmanned aerial vehicles \cite{Wang2022}. As a distributed and collaborative machine learning (ML) approach, FL enables users to have indirect access to a larger dataset while preserving privacy. Specifically, the devices train local ML models and transmit periodic updates to a base station (BS). The BS then aggregates these local updates into a global model and transmits it to the devices. This process is repeated over multiple communication rounds until the global model converges. 

The iterative nature of FL and the substantial size of local updates raise concerns about spectral and energy efficiency. Note that many IoT devices rely on limited battery capacity, making continuous network and FL operation a key challenge. Also, energy harvesting (EH) has emerged as a solution for recharging batteries externally or to avoid their replacement altogether~\cite{Lopez2021, Lopez2023}, but the variability and unpredictability of ambient energy sources demand careful consideration~\cite{Lopez2024}.  To address these challenges, new protocols are needed to enable the reliable execution of FL, where it is essential to balance the collaborative benefits of information sharing inherent to FL with the specific demands of \mbox{EH-IoT} networks.

Federated distillation (FD)~\cite{Jeong2018, Oh2020} aims to address the communication overhead and energy challenges in FL by sharing model logits, which are typically much smaller than the entire model. Collaborative learning is achieved through knowledge distillation (KD), in which a regularization term is integrated into the local loss function to distill knowledge from teacher to student. The regularization term measures the difference between the local model logits (student model) and the global logits (teacher model). Although FD reduces both energy consumption and communication load with a relatively simple procedure, this comes at the cost of reduced model accuracy~\cite{Seo2020}, especially when users do not have independent and identically distributed (IID) datasets. These settings are highly prevalent in IoT systems, where user data often vary significantly between devices due to differences in user behavior, geographic location, or data collection environments. 

\subsection{Related Work}
Existing FL literature often addresses energy and communication efficiency concerns by optimizing convergence time, reducing the required number of iterations, and the corresponding communication costs. In~\cite{Chen2021}, the authors propose a probabilistic user selection scheme combined with artificial neural networks to estimate the local models of non-selected users, thereby improving the convergence time. However, this approach requires one user to maintain constant communication with the BS throughout all training iterations. In~\cite{Pase2021}, the convergence time is optimized by restricting users that cannot meet a predefined transmission rate from sending local model updates. The method relies on imperfect channel state information, which can be difficult to estimate for IoT devices. The work in~\cite{Luo2024} introduces a client sampling method designed to handle heterogeneous settings while incorporating adaptive bandwidth control. In~\cite{Wu2022}, the relationship between local and global models is used to identify and exclude local updates that adversely contribute to the global model. The method proposed in~\cite{Zhou2022} improves spectral efficiency by parallelizing the model training and communication phases, overlapping the transmission of local model updates with the global model download, thereby reducing overall engagement time. Moreover, a problem to minimize the total training time is formulated in~\cite{Zhang2022} considering both the per-round latency and the total number of rounds. Although the above strategies improve the convergence time, they overlook the increased protocol complexity that limits their applicability in IoT networks.

Motivated by the growing need for decentralized learning in dynamic IoT networks, it becomes critical to investigate how FL can be effectively integrated with random access (RA) protocols. RA, which manages uncoordinated and decentralized access to shared wireless channels, underpins the communication framework of typical EH-IoT networks due to its simplicity and scalability~\cite{Clazzer2019}. Recognizing these advantages, recent works have already begun exploring the deployment of FL in RA communication settings~\cite{Xia2024, Evgenidis2024}. For example, sporadic user participation is considered in \cite{Choi2020}. The authors show that slotted ALOHA (SA) with multiple channels can perform better than sequential polling, as idle channels are mitigated. The work in~\cite{Valente2023} extends \cite{Choi2020} to EH-IoT networks and introduces an energy-aware random sleep probability to balance the availability and expenditure of energy resources, achieving energy neutrality. This method mitigates battery depletion and improves overall accuracy in critical EH scenarios. Furthermore, the works \cite{Valente2022, Salama2023} also consider RA for FL. The former applies sparsification in an FL scenario with multiple models, followed by channel allocation to achieve multi-model convergence. The latter shows that communication costs can be reduced by adopting SA in networks with mesh topology, indicating greater flexibility for IoT systems. 

Although the above works consider and often highlight the benefits of simple RA protocols in IoT networks, they do not account for the fact that many IoT applications operate on shared wireless channels. In such settings, unrelated background activity can significantly affect performance. This is partially addressed in~\cite{Mahmoudi2020_ICC}, wherein the latency of an SA network with one replica is optimized in a distributed learning context. However, such an approach does not address the interference caused by high background traffic environments, which hinders the participation of active users. This problem is accentuated when considering the prohibitively large size of local updates in FL.

Motivated by the high energy consumption and spectrum inefficiency associated with transmitting large local model updates in FL, FD algorithms have gained significant attention in recent years despite the loss of accuracy performance caused by data heterogeneity. For example, the work in~\cite{Ahn2019} focuses on improving performance considering a hybrid approach between FD and independent learning. Note that FD is derived from the concept of KD, wherein, both the teacher and the student require access to the same training data. Thus, to improve the accuracy of FD algorithms, a common strategy is to use a public dataset. In~\cite{Itahara2023}, the authors use a public unlabeled dataset to further train an FD framework and improve its performance via data augmentation while maintaining a reduced communication overhead compared to FL. The local model outputs exchanged during the FD iteration are used to label the data from the open dataset. Moreover, to prevent ambiguous labeling caused by heterogeneity in the data, an entropy-reducing average is proposed in the model output aggregation. Similarly, FedMD~\cite{Li2019} and FedED~\cite{Sui2020} use public datasets as a form of data augmentation to increase FD accuracy. FedMD trains each model in a public dataset before fine-tuning it in a private dataset. Meanwhile, FedED proposes that the student models learn locally and use a shared dataset to generate predictions, which are uploaded to a server to create a virtual teacher. Thus, the server can calculate the distillation loss between the teacher and the students. In addition, an auxiliary dataset is used in~\cite{Shao2024} to improve knowledge sharing in FD by filtering out misleading knowledge from atypical samples in the upload phase. 

However, the above approaches require a public dataset, which could be limited in size due to privacy issues and not scale in the same proportion as IoT networks. In addition, delivering these datasets to IoT devices can be challenging due to network capacity and memory limitations. Alternatively, synthetic datasets are also used in FD \cite{Jeong2018, Zhenyuan_Zhang2022, Zhou2023}. In \cite{Jeong2018}, each device collectively trains a generative model for data augmentation to create an IID dataset. Similarly, the use of generative models is proposed in~\cite{Zhenyuan_Zhang2022, Zhou2023} to address the non-IID nature of local datasets. Specifically, the work in~\cite{Zhenyuan_Zhang2022} considers a three-player generative adversarial network (GAN), while a GAN is employed in~\cite{Zhou2023} in the context of digital twins. However, a key limitation of synthetic data is the need for accurate knowledge of the true data distribution, which is often unknown. This can result in synthetic data that poorly represents the actual dataset and introduces inaccuracies rather than reducing them.

\subsection{Contributions}
In this paper, we investigate the performance of FD and FL networks under the typical energy and communication resource limitations of EH-IoT networks, and propose a new approach to jointly exploit their distinctive advantages. For this, we adopt a multi-channel SA protocol, which is well suited for managing the varying access patterns typical of EH-IoT scenarios. We incorporate background traffic to better reflect realistic network conditions.

The proposed method aims to improve resource efficiency and convergence time compared to traditional FL, while enhancing the model accuracy in comparison to both FL and FD. To support these objectives, it employs a low-complexity protocol that does not rely on a proxy dataset or the transmission of linearly mixed local samples to the server, as done in Mix2FLD~\cite{Oh2020}. The main contributions of the paper are summarized as follows.

\begin{enumerate}
    \item We formulate the uplink throughput of a multichannel SA system, defined as the rate at which local updates are successfully delivered to the server. Our formulation considers error correction codes, background traffic activity following a Poisson distribution, and the use of multiple subpackets. This formulation addresses scenarios where the size of local updates in FL exceeds the transmission capacity of a single packet, a common limitation of conventional IoT technologies. We then discuss how the background traffic impacts the uplink throughput and the entailed implications for FL and FD performance.
    \item We propose FL-distillation alternation (FLDA), where users alternate between FD and FL phases to leverage the strengths of both approaches. Initially, FD is performed, followed by a transition to FL. However, due to the limited information exchanged in FD, users' local models can diverge in non-IID settings, which causes a consensus phase when switching to FL. To address this, our method implements more frequent and multiple switches between FD and FL, ensuring that each user's local model remains aligned with a more up-to-date global model.
    \item We assess the performance of FLDA against FL and FD via simulations. We show that the proposed method outperforms FD in terms of accuracy, and FL in both convergence time and accuracy while using less energy to reach target accuracy goals. For example, in simulations using the MNIST dataset~\cite{MNIST2010} in a non-IID setting with a six-layer convolution neural network (CNN) model, FLDA uses $51$\% less energy than FL to reach $80$\% accuracy, while FD cannot reach the target accuracy. When considering lower accuracy levels, FLDA can save up to $98$\% energy compared to FL. Moreover, the introduced scheme copes better with the adversities of EH-IoT scenarios, being less sensitive to background traffic activity and critical EH conditions.
\end{enumerate}

The remainder of this paper is organized as follows. In Section~\ref{sec:preliminares}, we present FL and FD preliminaries, detailing how these approaches work. Section~\ref{sec:system_model} introduces the communication, energy consumption, and EH models. In Section~\ref{sec:proposed_scheme}, we formalize the proposed method and formulate the analytical uplink throughput of the system. Section~\ref{sec:numerical_evaluation} shows the performance of the outlined strategy compared to FL and FD, and Section~\ref{sec:conclusion} concludes the paper. In Table~\ref{tab:symbols_list}, we present the symbols used in this paper with their meanings.
\begin{table}[h]
\centering
\caption{List of symbols.}
\scriptsize
\addtolength{\tabcolsep}{-0.4em}
\begin{tabular}{l|l}
\midrule
\textbf{Symbol} & \textbf{Meaning}\\ \midrule
$\alpha$ & Ratio of iterations dedicated to FD \\
$B_{\text{max}}$ & Battery capacity of the EH-IoT devices\\
$B(z, t)$ & Instantaneous battery of the EH-IoT devices \\
$\beta(t)$ & Regularization gain at the $t$-th iteration\\ 
$\mathcal{B}_k(t)$ & Mini-batch dataset at the $t$-th iteration\\ 
$C$ & Number of labels \\
$D$ & Number of information subpackets during uplink\\
$\mathcal{D}_k$ & Local dataset  \\
$E^\text{cmp}_k$ & Energy required for the computation phase \\ 
$E^{\text{comms}}_{k, \text{FD}}$, $E^{\text{comms}}_{k, \text{FL}}$ & FD/FL communication energy consumption \\ 
$E^{\text{tx}}_{k, \text{FD}}$, $E^{\text{tx}}_{k, \text{FL}}$ & FD/FL transmission energy consumption \\ 
$E^{\text{rx}}_{k, \text{FD}}$, $E^{\text{rx}}_{k, \text{FL}}$ & FD/FL reception energy consumption \\ 
$\eta$ & Transceiver power amplifier efficiency \\ 
$f_k(\cdot)$ & Local loss function \\ 
$f_{\text{clk}, k}$ & Processor clock frequency \\ 
$F \in \{F_{\text{FD}}$, $F_{\text{FL}}\}$ & Number of subpacket transmissions for FD/FL\\
$\gamma$ & Number of iterations in a full cycle \\
$\mathbf{g}_{k, \text{FD}}(t)$, $\mathbf{g}_{k, \text{FL}}(t)$ & FD/FL gradient \\ 
$G_k$ & FLOPs processed in one iteration \\ 
$\mathbf{G}_n(t)$ & Global average (avg.) output vector \\
$K$ & Number of EH-IoT devices \\
$\ell^j_{n}$ & $n$-th label in the $j$-th label vector \\ 
$\mathbf{L}^i_{k, n}(t)$ & Local output vector \\ 
$\Bar{\mathbf{L}}_{k, n}(t)$ & Avg. output vector for label $\ell_n$ \\ 
$\Bar{\mathbf{L}}_k(t)$ & FD local update \\ 
$\lambda$ & Avg. background load \\
$M$ & Number of channels \\
$\mathcal{M}_{k, n}(t)$ & Samples with label $\ell_n$ as ground truth \\ 
$m_{k, n}(t)$ & Size of $\mathcal{M}_{k, n}(t)$ \\ 
$\mu$ & Learning rate \\
$N \in \{N_{\text{FD}}, N_{\text{FL}}\}$ & local update size for FD/FL \\
$N_s$ & Maximum BLE payload \\
$\varphi(\cdot)$ & Regularization function \\ 
$P^\text{cmp}_k$ & CPU power consumption \\ 
$P_{\text{circ}}$ & Transceiver fixed power consumption \\ 
$P^{\text{total}}_k$ & Transmission power consumption \\ 
$P_k^{\text{tx}}$ & Radiated transmission power \\ 
$P_k^{\text{rx}}$ & Transceiver reception power \\ 
$p_k$ & Access probability of the $k$-th user \\ 
$\psi_k$ & Effective capacitance of the CMOS chip \\ 
$q$ & Error correction code rate \\
$R^{\text{rx}}_b$, $R^{\text{tx}}_b$ & Downlink/uplink bit rates \\
$r$ & Energy income rate \\
$\varrho$ & Avg. energy income \\ 
$\rho_{\text{FD}}$, $\rho_{\text{FL}}$, $\rho_{\text{FLDA}}$ & Packet uplink throughput with FD/FL/FLDA \\ 
$\tau_s$ & Subpacket transmission time \\
$\Theta_k$ & FLOPs processed per cycle \\ 
$W$ & Number of FLOPs per data sample \\ 
$\mathbf{w}(t)$ & FL global update \\ 
$\mathbf{w}_k(t)$ & Local model of the $k$-th user \\ 
$\mathbf{x}^j_k$, $\mathbf{y}^j_k$ & $j$-th unlabeled sample and label vector \\ 
\midrule
\end{tabular}
\label{tab:symbols_list}
\end{table}

\section{Preliminaries}
\label{sec:preliminares}
Consider $K$ FL clients, indexed as $k \in  \mathcal{K} = \{1, 2, \ldots, K\}$, and a central server functioning as an aggregator. Each client has a local dataset \mbox{$\mathcal{D}_k = \{(\mathbf{x}^{(j)}_k, \mathbf{y}^{(j)}_k)\}$} associated with its local model, where $\mathbf{x}^{(j)}_k$ represents the $j$-th unlabeled sample, and \mbox{$\mathbf{y}^{(j)}_k = \{\ell_n^{(j)}\}_{n=1}^{C}$} denotes the label vector used for supervised learning in a classification task with $C$ labels. The element $\ell^{(j)}_{n}$ equals to $1$ if the \mbox{$n$-th} label is the ground truth and we have $0$ otherwise. Each client aims to minimize its local loss function, which is defined differently for the FL and FD cases.

\subsection{Federated Learning}
\label{subsec:federated_learning}

\begin{figure*}[t!]
    \centering
    \subfloat[]{%
        \includegraphics[width=0.43\textwidth, trim=0mm 0mm 0mm 0mm, clip]{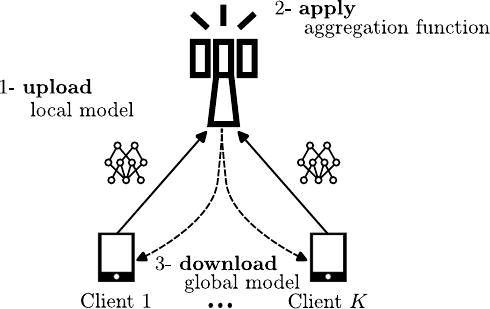}%
        \label{fig:FL_iteration}
    }
    \hfill % Adds horizontal space between the subfigures
    \subfloat[]{%
        \includegraphics[width=0.48\textwidth, trim=0mm 0mm 0mm 0mm, clip]{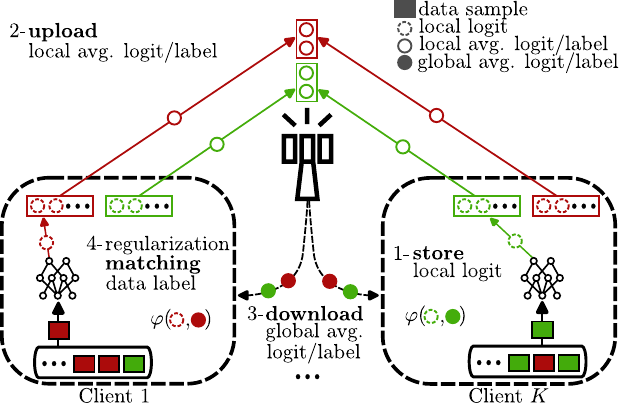}%
        \label{fig:FD_iteration}
    }
    \caption{Distributed learning procedures: (a) FL iteration, (b) FD with $2$ possible labels.}
    \label{fig:iteration_procedures}
\end{figure*}
FL aims to optimize the model weights to minimize the global loss function, i.e.,
\begin{subequations}\label{eq:problem_global_loss_function}
\begin{align}
        \label{eq:problem_global_loss_function_a}
        &\minimize_{\mathbf{w}, \{\mathbf{w}_k\}_{k=1}^K}\quad\dfrac{1}{K}\sum\limits_{k \in \mathcal{K}} f_k(\mathbf{w}_k, \mathcal{D}_k) \\
        &\textrm{subject to}\quad\mathbf{w}_k = \mathbf{w}, \quad\forall k, \label{eq:problem_global_loss_function_constraint}
\end{align}
\end{subequations}\unskip where \mbox{$f_k(\mathbf{w}_k, \mathcal{D}_k)$} is the local loss function associated with the $k$-th client, $\textbf{w}_k$ are the local model parameters, and $\mathbf{w}$ is the global model. The consensus constraint in~\eqref{eq:problem_global_loss_function_constraint} ensures that all local models remain aligned with the global model.

In FL, clients do not upload raw data to the aggregator, but iteratively solve the problem in \eqref{eq:problem_global_loss_function} by uploading their local updates, i.e., local models, as shown in Fig.~\ref{fig:FL_iteration}. Specifically, the minimization process is carried out using stochastic gradient descent (SGD), with the gradient defined as
\begin{equation}
    \mathbf{g}_{k, \text{FL}}(t) = \dfrac{1}{|\mathcal{B}_k(t)|}\sum\limits_{(\mathbf{x}_k, \mathbf{y}_k) \in \mathcal{B}_k(t)} \nabla f_k(\mathbf{w}_k(t), \mathbf{x}_k, \mathbf{y}_k),
    \label{eq:local_update_FL}
\end{equation}
where \mbox{$\mathcal{B}_k(t) \subseteq \mathcal{D}_k$} is the local mini-batch dataset. After computing the gradient, each client updates its model by taking a minimization step in the opposite direction to the gradient, i.e., 
\begin{equation}
    \mathbf{w}_k(t+1) = \mathbf{w}_k(t) - \mu\mathbf{g}_{k, \text{FL}}(t),
    \label{eq:optimization_step_FL}
\end{equation}
where \mbox{$\mu > 0$} is the learning rate.

To ensure collaborative learning, i.e, leveraging the local datasets of all clients, each client uploads its local model to the central server. The central server then aggregates the local models to create the global model, which is given by
\begin{equation}
    \mathbf{w}(t+1) = \sum\limits_{k \in \mathcal{K}} b_k\mathbf{w}_{k}(t+1).
    \label{eq:aggregation_FL}
\end{equation}
Here, \mbox{${b_k = |\mathcal{B}_k(t)|/\sum_{k^{'}=1}^K |\mathcal{B}_k^{'}(t)|}$} captures the relative size of the mini-batch. To finalize the iteration, the central server broadcasts the global model and the clients reinitialize their weights at the next iteration as
\begin{equation}
    \mathbf{w}_k(t+1) = \mathbf{w}(t+1).
    \label{eq:reinitialization_FL}
\end{equation}
This process repeats until the problem in \eqref{eq:problem_global_loss_function} is solved with an arbitrary low value of the loss function in~\eqref{eq:problem_global_loss_function_a}.

\subsection{Federated Distillation}
\label{subsec:federated_distillation}

Unlike FL, FD focuses on transmitting local averaged output vectors to the aggregator, separately for each ground truth label, since these are usually much smaller than the whole model. This is illustrated in the FD iteration procedure in Fig.~\ref{fig:FD_iteration}. 

Let us assume that \mbox{$\mathcal{M}_{k, n}(t) \subseteq \mathcal{B}_k(t)$} is the subset of samples with the label $\ell_n$ as the ground truth. Here, \mbox{$|\mathcal{M}_{k, n}(t)| = m_{k, n}(t)$} and \mbox{$\sum_n m_{k, n}(t) = |\mathcal{B}_k(t)|$}. Thus, the local averaged output vectors are given by 
\begin{equation}
    \Bar{\mathbf{L}}_{k, n}(t) = \dfrac{1}{m_{k,n}(t)}\sum\limits_{i = 1}^{m_{k,n}(t)} \mathbf{L}^{(i)}_{k, n}(t),
    \label{eq:local_update_FD}
\end{equation}
where $\mathbf{L}^{(i)}_{k, n}(t)$ is the local output vector, i.e., softmax logits, for the $i$-th sample of subset $\mathcal{M}_{k, n}(t)$. Since there are $C$ possible labels, each client transmits $C$ local averaged output vectors, \mbox{$\Bar{\mathbf{L}}_k(t) = \{\Bar{\mathbf{L}}_{k, 1}(t), \ldots, \Bar{\mathbf{L}}_{k, C}(t)\}$}. The central server averages $\Bar{\mathbf{L}}_{k, n}(t)$ across clients, resulting in the global averaged output vectors for each possible label 
\begin{align}
    \mathbf{G}_n(t) = \dfrac{1}{K}\sum_{k \in \mathcal{K}} \Bar{\mathbf{L}}_{k, n}(t).
    \label{eq:aggregation_FD}
\end{align}
Then $\mathbf{G}(t) = \{\mathbf{G}_1(t), \ldots, \mathbf{G}_{C}(t)\}$ is broadcast to the clients, which use it to regularize the cost function, penalizing the gap between the local output vector $\mathbf{L}^{(i)}_{k, n}(t)$ and the global averaged output vector $\mathbf{G}_n(t)$. This promotes collaborative learning in the form of KD. Therefore, the gradient of local loss function for FD is given by
\begin{align}
    \mathbf{g}_{k, \text{FD}}(t) = \dfrac{1}{|\mathcal{B}_k(t)|}
    &\sum\limits_{n=1}^{C} \sum\limits_{i = 1}^{m_{k,n}(t)} \Bigg[ \nabla f_k\left( \mathbf{w}_k(t), \mathbf{x}^{(i)}_{k}, \mathbf{y}^{(i)}_k \right) \nonumber \\
    &+ \beta(t) \nabla \varphi\left( \mathbf{L}_{k, n}^{(i)}(t), \mathbf{G}_n(t) \right) \Bigg],
    \label{eq:FD_gradient}
\end{align}
where $\varphi(\cdot)$ is the regularization function, and $\beta(t)$ is a constant that defines the intensity of the regularization, which is set to \mbox{$\beta(t) = 0$} if the client could not receive $\mathbf{G}(t)$ at the $t$-th iteration. Clients then reinitialize their weights in the next iteration as
\begin{equation}
    \mathbf{w}_k(t+1) = \mathbf{w}_k(t) - \mu\mathbf{g}_{k, \text{FD}}(t).
    \label{eq:reinitialization_FD}
\end{equation}

\section{System Model}
\label{sec:system_model}
\begin{figure}[t!] 
    \centering
    \includegraphics[width=0.5\textwidth, trim=0mm 0mm 0mm 0mm, clip]{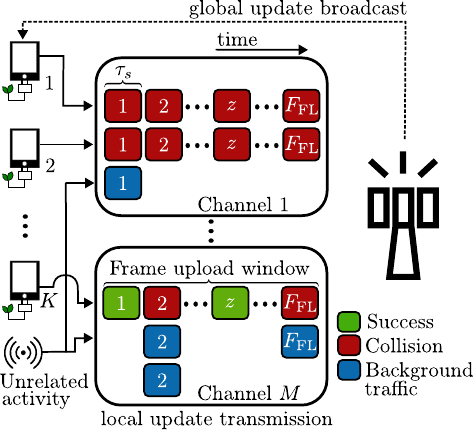}%
    \caption{Communication system model. The illustrated scenario depicts devices performing FL, resulting in a frame upload window of $\tau_s F_{\text{FL}}$. If FD were employed instead, the frame upload window would be $\tau_s F_{\text{FD}}$, reflecting the smaller size of FD updates.}
    \label{fig:communication_system_model}
\end{figure}

The practical execution of both FL and FD processes, described in the previous section, depends on communication with the central server and the energy availability of clients. To capture these factors, this section introduces the system model and details the operational conditions that influence the performance of these learning strategies within an EH-IoT context.

\subsection{Communication Model}
Consider a wireless network consisting of $K$ EH-IoT users, each functioning as an FL client with finite battery capacity $B_{\text{max}}$. Users communicate with a BS, which acts as the central server, over $M$ orthogonal channels. Additionally, the local update size in FL may be too large for users to transmit in one packet (i.e., without data splitting into several subpackets) with conventional IoT technologies. Thus, we consider that a transmission of a subpacket takes $\tau_s$ seconds with an uplink bit rate of $R^{\text{tx}}_b$ bits/s, resulting in a payload size of $N_s = \tau_sR^{\text{tx}}_b$. 

The number of subpackets required to complete the transmission of a full update is given by
\begin{equation}
    F =  \left\lceil D/q \right\rceil
\end{equation}
where $D=\left\lceil {N}/{N_s} \right\rceil$ is the number of information subpackets and $N$ represents the local update size, which is set to $N_{\text{FL}}$ or $N_{\text{FD}}$ for FL and FD, respectively. Moreover, \mbox{$q \in (0,1]$} is the error correction code rate, included for fairness purposes. Its value indicates the proportion of a packet allocated to information bits, while the remaining $1-q$ is used for redundancy (error correction purposes).

Because the local updates for FL and FD differ in size, the total transmission duration for each update also varies. As shown in Fig.~\ref{fig:communication_system_model}, we refer to this duration as the frame upload window, that is, the time required to transmit a packet. Thus, $F_{\text{FL}}$ also represents the number of discrete time intervals of $\tau_s$ seconds needed for the transmission of the FL local update, while for FD it is $F_{\text{FD}}$. Throughout this paper, we define an interval of $\tau_s$ seconds as a time slot.

Given that users share access to the wireless medium, simultaneous transmissions by multiple users may result in collisions. In this work, we assume a hard collision model, where any overlap in transmissions causes all involved packets to fail. We also consider uncorrelated background traffic to reflect realistic wireless network conditions where users share the medium with external, unrelated transmissions. Here, the background traffic is modeled as a Poisson random variable \mbox{$\upsilon(z, t) \sim \mathrm{Poisson(}\lambda\mathrm{)}$}, where~$z$ represents the $z$-th subpacket and $\lambda$ is the background traffic load. That is, $\lambda$ is the average number of background subpackets that are in the network during each time slot, indexed by $z$. Specifically, for each iteration~$t$, $z$ spans from~$1$ to the total number of subpackets in the frame window and resets to~$1$ at the start of each new $t$, which is always rising in discrete steps. Note that this model generalizes to the ideal scenario without background activity when $\lambda = 0$.

An iteration, illustrated in Fig.~\ref{fig:communication_system_model}, consists of the following $5$ steps:
\begin{enumerate}
    \item \textbf{Energy Harvesting:} At the start of the \mbox{$z$-th} time slot, the $k$-th device harvests \mbox{$\zeta_k(z, t)$} Joules of energy and stores in the battery if its capacity allows. We assume \mbox{$\zeta_k(z, t)$}~is a random variable with a predefined distribution as described in Section~\ref{subsec:energy_harvesting_model}.
    
    \item \textbf{Computation:} At the start of each new iteration, users compute their local model based on \eqref{eq:local_update_FL} or \eqref{eq:local_update_FD}, depending on the adopted learning system. This step only occurs at the beginning of an iteration, i.e., $z = 1$, as local models are computed once per iteration.
    
    \item \textbf{Channel Access:} Each device transmits the $z$-th subpacket of the $t$-th local update with probability \mbox{$p_k$}. Transmissions occur through a randomly chosen channel among $M$ channels. Simultaneously, \mbox{$\upsilon(z, t)$} background subpackets are transmitted, each one over any of the $M$ channels with a uniform probability. The transmission of a user subpacket is only successful if there is no collision, while a packet is successfully received if \mbox{$\lceil qF_{\text{FL}} \rceil$} \mbox{($\lceil qF_{\text{FD}} \rceil$)} subpackets reach the BS without collisions in the case of FL (FD).
    
    \item \textbf{Global model updates:} The BS aggregates local updates that were successfully received into a global model upon the closing of a frame window. This happens by following \eqref{eq:aggregation_FL} for FL and \eqref{eq:aggregation_FD} for FD. Then, the BS broadcasts the global update back to the users.
    
    \item \textbf{Model re-initialization:} Following \eqref{eq:reinitialization_FL} or \eqref{eq:reinitialization_FD}, the users reinitialize their local model weights and the next iteration begins. 
\end{enumerate}
Note that steps $1$) and $3$) occur in every time slot, while steps $2$), $4$), and $5$) occur at the beginning or end of an iteration. Following these steps, we model the battery state evolution as
\begin{align}
    B_k(z, t) &= B_k(z-1, t) + \min(\zeta_k(z, t), B_\text{max} - B_k(z-1, t)) \notag \\
    &\quad - \delta_{\text{tx}, k}(z, t) \tau_s P^{\text{total}}_k - \delta_{\text{rx}, k}(z, t) E^\text{rx}_k \notag \\
    &\quad - \delta_{\text{cmp}, k}(z, t) E^\text{cmp}_k,
\end{align}
where $\delta_{\text{tx}, k}(z, t)$, $\delta_{\text{rx}, k}(z, t)$, and $\delta_{\text{cmp}, k}(z, t)$ are indicator functions representing the occurrence (or not) of a transmission, reception, and computation phase, respectively. That is, they are equal to~$1$ if the corresponding event occurs and equal to~$0$ otherwise. An event might not occur because of battery depletion or, in the case of computation and reception, if the transmission of a packet is still ongoing. Furthermore, $E^\text{rx}_k$ and $E^\text{cmp}_k$ are the computation and reception energy costs, while~$P^{\text{total}}_k$~is the power consumption during transmission, which are modeled next.

\subsection{Local-Computation Energy Consumption Model}
The computation complexity of a ML algorithm can be measured by the number of required floating point operations (FLOPs). Let $W$ denote the number of FLOPs per data sample for a given model. The total number of FLOPs for the $k$-th user to perform one local update is
\begin{equation}
 G_k = W|\mathcal{B}_k|.
\label{eq:Flops_update}
\end{equation}
Let $f_{\text{clk}, k}$ be the processor clock frequency (in cycles/s) of the $k$-th user and $\Theta_k$ be the number of FLOPs it processes within one cycle. Then, the time required for one local update is
\begin{equation}
    t_k = \dfrac{G_k}{\Theta_kf_{\text{clk}, k}}, \quad \forall k \in \mathcal{K}.
    \label{eq:time_localupdate}
\end{equation}
Moreover, for a CMOS circuit, the central processing unit (CPU) power is often modeled by its most predominant part: the dynamic power \cite{Zhang2013}, which is proportional to the square of the supply voltage and to the operating clock frequency. Moreover, for a low voltage supply, as in our case, the frequency scales approximately linear with the voltage \cite{Zhang2013}. Therefore, the CPU power consumption can be written as \cite{Zeng2022}
\begin{equation}
    P^\text{cmp}_k = \psi_k f_{\text{clk}, k}^3 \quad \forall k \in \mathcal{K},
    \label{eq:power_CMOS}
\end{equation}
where $\psi$ is the effective capacitance and depends on the chip architecture. Based on \eqref{eq:time_localupdate} and \eqref{eq:power_CMOS}, the energy consumption of the computation phase for the $k$-th user is given by
\begin{equation}
    E^\text{cmp}_k = t_k P^\text{cmp}_k = \psi_k \dfrac{G_k}{\Theta_k}f_{\text{clk}, k}^2.
    \label{eq:Energy_cmp}
\end{equation}

Note that the only difference in the computation phase between FL and FD is the local regularization function in~\eqref{eq:FD_gradient}. Since the size of $\mathbf{w}_k(t)$ is much larger than that of $\mathbf{L}^{(i)}_{k,  n}(t)$, the total number of FLOPs $G_k$ does not change significantly, resulting in an insignificant impact on the overall energy consumption. Therefore, the energy cost in the computation phase is assumed to be the same for both methods.

\subsection{Transceiver Energy Consumption Model}
The energy consumed by the IoT transceivers is 
\begin{equation}
    E^{\text{comms}}_k = E^{\text{tx}}_k + E^{\text{rx}}_k + E^{\text{sleep}}_k,
    \label{eq:energy_comms}
\end{equation}
where $ E^{\text{tx}}_k$ ($E^{\text{rx}}_k$) is the energy required to transmit (receive) a local (global) update while $E^{\text{sleep}}_k$ is the energy consumed during the inactive time. Since $E^{\text{sleep}}_k$ is much smaller than $E^{\text{tx}}_k$ and $E^{\text{rx}}_k$, we neglect its impact in the following. 

Considering the transmission of local updates with a radiated power $P_k^{\text{tx}}$, the power consumed by the IoT device can be modeled as \cite{Scaciota2022}
\begin{equation}
    P^{\text{total}}_k = \dfrac{P^{\text{tx}}_k}{\eta} + P_{\text{circ}},
    \label{eq:ptx}
\end{equation}
where $\eta$ is the drain efficiency of the power amplifier (PA), and $P_{\text{circ}}$ is a fixed power consumption that comprises all other transceiver circuits except the PA. Then, the energy required to transmit a local update is given by
\begin{align}
    E^{\text{tx}}_{k} = \dfrac{P^{\text{total}}_k}{R^{\text{tx}}_b}N.
\end{align}
Here, if $N = N_{\text{FL}}$, the energy required to transmit a local update is $E^{\text{tx}}_{k, \text{FL}}$. Similarly for $N = N_{\text{FD}}$ in the FD case.
Meanwhile, the energy consumed when receiving the global updates is generally modeled by
\begin{equation}
    E^{\text{rx}}_k = \dfrac{P^{\text{rx}}_k}{R^{\text{rx}}_b}N,
\end{equation}
where $R^{\text{rx}}_b$ is the bit rate in the downlink, and $P^{\text{rx}}_k$ is the power consumed during reception, accounting for the operation of the receiver’s circuitry. Moreover, $N \in \{N_\text{FL}, N_\text{FD}\}$ represents the global update size in bits. Depending on the method each user applies, the energy consumption to receive an update is $E^{\text{rx}}_{k, \text{FL}}$ or $E^{\text{rx}}_{k, \text{FD}}$ for FL and FD, respectively.

Note that $N_{\text{FL}}$ is much larger than $N_{\text{FD}}$, indicating that FL may require significantly more energy than FD to complete the transmission of a local update and to receive the global model.

\subsection{Energy Harvesting Model}
\label{subsec:energy_harvesting_model}
Similar to~\cite{Hamdi2022}, we adopt a compound Poisson process to model the energy income at each user. That is, the interarrival time is modeled by an exponential distribution with rate $r$, i.e., $r$ times every \mbox{$\tau_s$ seconds}, and the amount of energy harvested in each arrival is modeled by a Poisson process with parameter \mbox{$\varrho/r$}, thus, \mbox{$\EX\left[\zeta_k(z, t)\right] = \varrho$}. This model is defined in discrete units of energy, and it is possible to scale one unit of energy to any amount in Joules. Since the analysis is performed on devices in critical energy income regimes, we consider \mbox{$\varrho \in [0, 1]$} and scale one unit of energy to \mbox{$(E^{\text{comms}}_{k, \text{FL}} + E^\text{cmp}_{k})/F_{\text{FL}}$}. Therefore, on average, a device that performs an FL iteration harvests $\varrho(E^{\text{comms}}_{k, \text{FL}} + E^\text{cmp}_{k})$ J of energy during its frame window. To maintain a consistent and fair comparison, we assume that the average harvested power remains the same regardless of the adopted learning system. Consequently, a device performing an FD iteration harvests \mbox{$\varrho(E^{\text{comms}}_{k, \text{FL}} + E^\text{cmp}_{k})F_{\text{FD}}/F_{\text{FL}}$} J on average during its frame window, where $F_{\text{FD}}/F_{\text{FL}}$ represents the ratio between the frame durations for FD and FL.

\section{Network Performance Metrics}
Building on the system model, it is important to characterize how communication dynamics, including collisions, background traffic, and energy constraints, influence user participation in the aggregation phase of the learning strategies. Therefore in this section we formulate network metrics that quantify user participation. In this regard, the main difference between the learning strategies is the size of the local update.

We begin by defining the success probability of a user in multichannel SA with $p_k = p$ for all $k \in \mathcal{K}$, expressed as the following binomial distribution
\begin{align} 
    p_a &= {M \choose 1} \frac {p}{M} \left ({1 - \frac {p}{M} }\right)^{\hat{K}-1} = p \left ({1 - \frac {p}{M} }\right)^{\hat{K}-1},
    \label{eq:pa}
\end{align}
where $p$ is the access probability, $1/M$ is the probability of the transmission occurring in a specific channel, and $\hat{K} \leq K$ is the number of active users in the network. This expression assumes only collision-based errors and a hard collision model, as well as a time slot that comprises all the information bits of a packet. Also, note that when $\hat{K} = K$, $p_a$ is maximized at $p = M/K$, as shown in \cite{Choi2020}. 

With \eqref{eq:pa}, we derive the success probability with Poisson background traffic, which is given by
\begin{align} 
    p_b &= p_a\sum\limits_{n = 0}^\infty \dfrac{\lambda^n e^{n}}{n!}\left(1 - \dfrac{1}{M}\right)^n = p_a e^{-\dfrac{\lambda}{M}} = p_a p_s,
\end{align}
where the expression inside the sum considers the probability that none of the $n$ background subpackets uses the same channel as the $k$-th user. Moreover, the infinity sum converges to $p_s =  e^{-\lambda/M}$. 

When considering $D$ information subpackets with an error correction code rate of $q$, the multichannel SA success probability, including the effect of background traffic, is given by
\begin{align}
    p_{\text{MA}} = p_a\sum\limits_{z=D}^{\lceil D/q \rceil} {\lceil 
D/q \rceil\choose z} p_s^{z}(1-p_s)^{\lceil D/q \rceil - z}.
\label{eq:pma}
\end{align}
Here, the BS decodes the packet if any combination of $D$ or more subpackets is received free from background traffic, which is captured by the expression inside the summation, and if a collision with users of the same network does not happen, what is captured by $p_a$. 

The average number of local updates successfully received by the BS at a given iteration is
\begin{equation}
    \rho = \hat{K}p_{\text{MA}}.
    \label{eq:throghput}
\end{equation}

The number of active users in \eqref{eq:throghput} is strongly influenced by the energy availability of each user. That is, a user can participate in an iteration if the following condition is met
\begin{align}
    B_k(0, t) + \sum\limits_{z=1}^{\lceil N/q \rceil} \zeta_k(z, t) \geq E^{\text{comms}}_k +E^{\text{cmp}}_k.
\end{align}
Therefore, the probability that a user is active is given by
\begin{equation}
    P_{\text{active}}(t) = F_\zeta\left(B_k(0, t) - E^{\text{comms}}_k -E^{\text{cmp}}_k\right),
\end{equation}
where \( F_\zeta(x) = \Pr\left(\sum_{z=1}^{\lceil N/q \rceil} \zeta_k(z, t) \leq x \right) \) is the cumulative distribution function of the total harvested energy over ${\lceil N/q \rceil}$ time slots. Therefore, the effective number of active users scales as 
\begin{align}
    \hat{K}(t) = KP_{\text{active}}(t),
\end{align}
which influences \eqref{eq:pa}, \eqref{eq:pma}, and consequentially the network uplink throughput in \eqref{eq:throghput}.

Furthermore, $p_{\text{MA}}$ depends on the number of information subpackets $D$, which are different for FL and FD. As a result, the behavior of $p_{\text{MA}}$ in response to background traffic depends on the adopted learning system. To achieve a balance in this regard, we propose a new strategy described in the next section, which allows greater information sharing than FD and increases the robustness to background traffic compared to FL.

\section{Federated Learning-Distillation Alternation}
\label{sec:proposed_scheme}
\begin{figure}[t!] 
    \centering
    \includegraphics[width=0.45\textwidth, trim=0mm 0mm 0mm 0mm,clip]{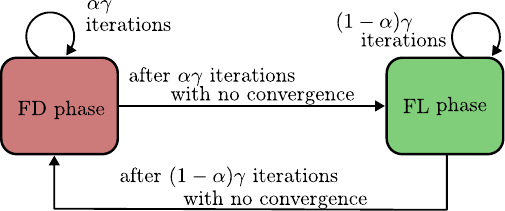} 
    \caption{Schematic of FLDA.}
    \label{fig:proposed_scheme_schematic}
\end{figure}
Considering the high accuracy potential of FL and the communication efficiency of FD, along with their respective limitations in energy consumption and accuracy under non-IID data, we propose a new scheme designed to leverage the strengths of both approaches. FLDA consists of periodic training cycles in which users alternate between FD and FL phases. A full cycle comprises $\gamma$ iterations, with each phase lasting $\alpha\gamma$ and $(1-\alpha)\gamma$ iteration for FD and FL, respectively, where $\alpha \in [0, 1]$ specifies the proportion of iterations dedicated to FD, and $1-\alpha$ the proportion dedicated to FL.

FLDA can operate effectively under varying network conditions, ensuring that knowledge exchange remains reliable despite interference and energy constraints. By alternating between FD and FL phases, the method balances communication efficiency with model accuracy, leveraging the effective uplink throughput of the network. Specifically, the FD phase provides a reliable means of knowledge sharing when FL transmissions experience losses due to interference, while the FL phase ensures that full-model updates are incorporated when network conditions are favorable. This behavior is indicated by considering the average uplink throughput $\rho_{\text{FLDA}}$ of FLDA over iterations, given as 
\begin{equation}
    \rho_\text{FLDA} = \alpha\rho_{\text{FD}} + (1-\alpha)\rho_{\text{FL}},
    \label{eq:throghput_hybrid}
\end{equation}
where $\rho_{\text{FL}}$ and $\rho_{\text{FD}}$ are the uplink throughput, defined in \eqref{eq:throghput}, using their corresponding local update sizes in bits.

The proposed FLDA scheme can be understood as an extension of FL with an adaptive regularization phase. Specifically, during the FD phase, users perform $\alpha\gamma$ local SGD steps with a regularization term applied to their loss function, allowing for the distillation of knowledge while reducing the communication overhead. However, unlike FL, users do not synchronize with the global model during this phase, which can lead to greater model divergence, particularly in non-IID settings. By switching to $(1-\alpha)\gamma$ iterations of FL, users reinforce the consistency of the global model, as they transmit the full updates of the local model and receive a new global model in this phase. This alternation balances communication costs and model consistency, and this compromise is influenced by the choice of parameters $\alpha$ and $\gamma$. High values of $\alpha\gamma$ might lead to a consensus phase when the switches occur from FD to FL. However, the opposite is not true, as the FL iterations serve to anchor each local model more closely to the global model. The alternation between FD and FL phases does not impose a strict optimization strategy, but rather accounts for challenging network conditions, ensuring that model updates are exchanged efficiently within the energy constraints of \mbox{EH-IoT} devices.

In terms of protocol complexity, FLDA introduces additional control for users to alternate between FD and FL phases. However, this can be implemented with minimal overhead. In practice, the BS can broadcast a phase identifier at the beginning of each cycle, indicating whether the upcoming iterations will follow FD or FL. The parameter $\alpha$ is pre-configured and broadcasted to all users during initialization. Users then internally count the number of iterations to ensure synchronization with the BS, eliminating the need for additional signaling during operation. This mechanism is closely aligned with traditional FL, where users already rely on the BS for synchronization and global model updates. Consequently, the added complexity is minimal, making the protocol feasible for \mbox{EH-IoT} networks. 

\section{Numerical Evaluation}
\label{sec:numerical_evaluation}
We analyze the performance of the proposed method compared to FL and FD. We set $\gamma = 100$ and $\alpha = 0.5$ unless otherwise indicated and train a six-layer CNN model\footnote{The CNN consists of two convolution layers with filter size $3 \times 3$ and ReLU activation. The first with $8$ feature maps and the second with $16$. Each filter is followed by a $2 \times 2$ max~pooling layer with stride $2$. The model then has a fully connected layer with $576$ units and ReLU activation, followed by the output layer with $10$ classes.} with the MNIST dataset, where \mbox{$|\mathcal{D}_k| = 500$}. Here, to construct a non-IID training dataset, each user has $2$ randomly selected labels with $2$ samples, and the other $8$ labels have $62$ samples. The test data set has \mbox{$5000$} sample images. We use the cross-entropy loss function for \mbox{$f_k(\mathbf{w}, \mathcal{D}_k)$} and \mbox{$\varphi( \mathbf{L}_{k, n}^{(i)}(t), \mathbf{G}_n(t) )$}~\cite{Oh2020}, and a maximum BLE payload size \mbox{$N_s = 2008$}~bits. Moreover, throughout this paper we use the values in Table~\ref{tab:default_values} unless stated otherwise.
\begin{table}[h]
\centering
\caption{Default simulation values}
\scriptsize
\renewcommand{\arraystretch}{1.2}
\begin{tabular}{l l | l l}
\toprule
\textbf{Variable} & \textbf{Value} & \textbf{Variable} & \textbf{Value} \\ 
\midrule
$\alpha$ & $0.5$ & $K$ & $20$ \\
$B_{\text{max}}$ & $0.1$ J & $M$ & $4$ \\
$\beta(t)$  & $1$ & $\mu$ & $0.01$ \\
$f_{\text{clk}, k}$ & $2$ GHz & $N_{\text{FD}}$ & $3200$ bits \\
$F_{\text{FD}}, F_{\text{FL}}$ & $4$, $220$ & $N_{\text{FL}}$ & $223488$ bits \\
$\gamma$ & $100$ & $N_s$ & $2008$ bits \\
$P_{\text{circ}}$ & $1.33$ mW~\cite{Tamura2020} & $P_k^{\text{tx}}$ & $3.3$ dB~\cite{Tamura2020} \\ 
$P_k^{\text{rx}}$ & $1.9$ mW~\cite{Tamura2020} & $p_k$ & $0.2$~\cite{Choi2020} \\ 
$\psi_k$ & $10^{-30}$~\cite{Zhao2022} & $q$ & $0.5$ \\
$R^{\text{rx}}_b$, $R^{\text{tx}}_b$ & $10^6$ bits/s & $r$ & $1/50$~\cite{Valente2023} \\
$\varrho$ & $0.4$ & $\Theta_k$ & $8$~\cite{Xu2021} \\ 
$W$ & $256896$~\cite{Mo2021} & & \\
\bottomrule
\end{tabular}
\label{tab:default_values}
\end{table}

\begin{figure}[t] 
    \centering
    \includegraphics[width=0.45\textwidth, trim=0mm 0mm 0mm 0mm,clip]{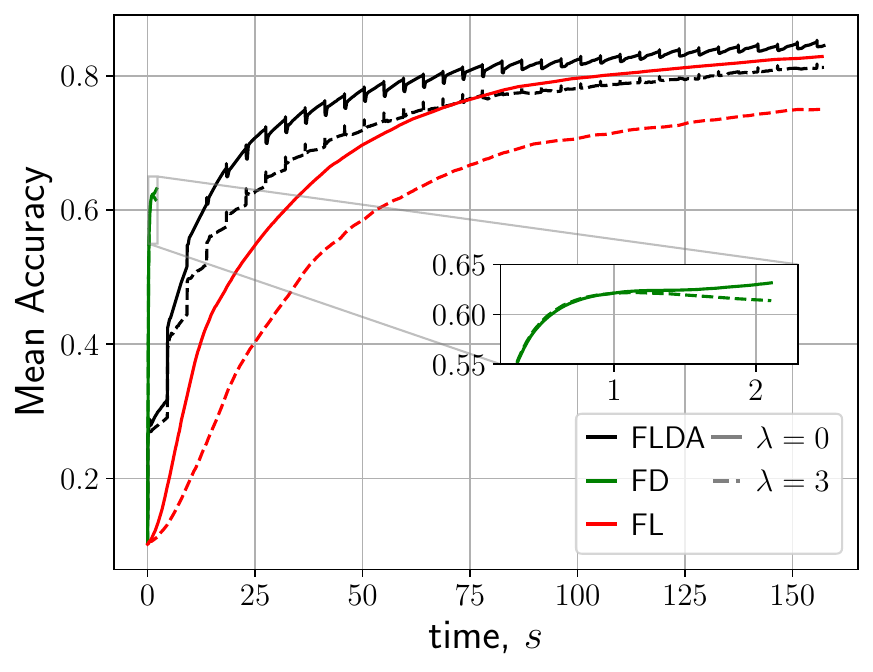} 
    \caption{Mean test accuracy as a function of time.}
    \label{fig:hybrid_accuracy_vs_time}
\end{figure}

First, we explore the mean accuracy behavior of the proposed approach as a function of time in Fig.~\ref{fig:hybrid_accuracy_vs_time} for a background traffic load $\lambda \in \{0, 3\}$. It is possible to see that the proposed approach has, to some extent, qualities of FD and FL. Specifically, it has a lower convergence time than FL and users share more information than FD on average, which helps it obtain a higher accuracy compared to FL and FD. Moreover, the developed approach performs FD for half of the iterations, making it less sensitive to high volumes of background traffic compared to FL. This can be seen by comparing the gap between the curves of the proposed method for $\lambda = 0$ and $\lambda = 3$ and that of the FL curves for the same values. Although FD consumes fewer communication resources and less energy per iteration, its final accuracy is significantly lower than that of the proposed method, highlighting the benefit of sporadically incorporating full model sharing.

\begin{figure}[t] 
    \centering
    \includegraphics[width=0.45\textwidth, trim=0mm 0mm 0mm 0mm,clip]{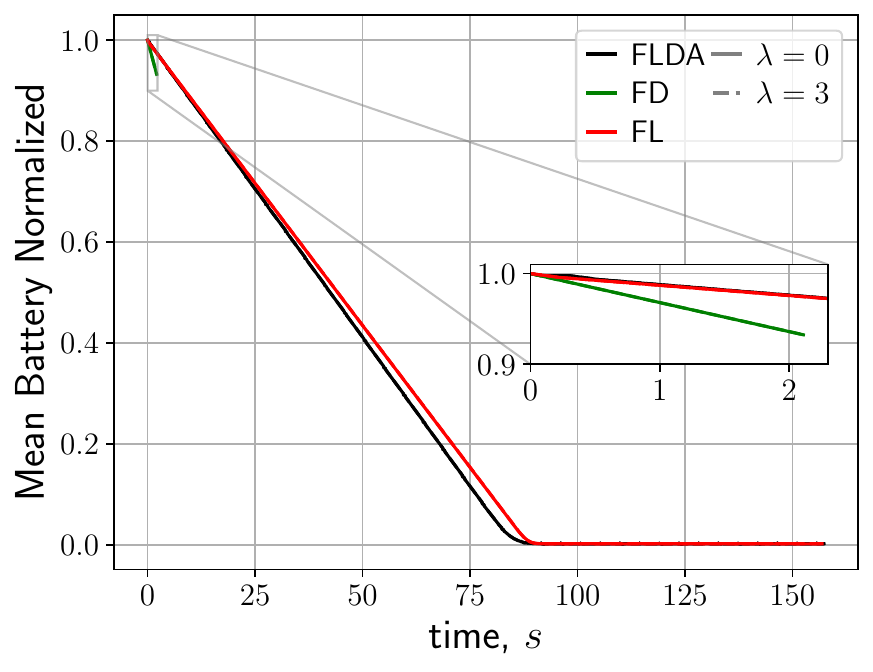} 
    \caption{Normalized average battery level as a function of time.}
    \label{fig:hybrid_battery_vs_time}
\end{figure}

Fig.~\ref{fig:hybrid_battery_vs_time} shows that the mean battery behavior over time for the proposed approach resembles more closely that of FL than FD. This is because the proposed approach takes more time to perform $0.5\gamma$ iterations of FL than the same amount of FD iterations. Notably, the battery level for the FD approach declines rapidly as multiple FD iterations occur within the time span of a single FL iteration, leading to additional computational costs. This also explains why the battery curve for the proposed method is slightly shifted to the left compared to the FL battery curve.

\begin{figure}[t] 
    \centering
    \includegraphics[width=0.45\textwidth, trim=0mm 0mm 0mm 0mm,clip]{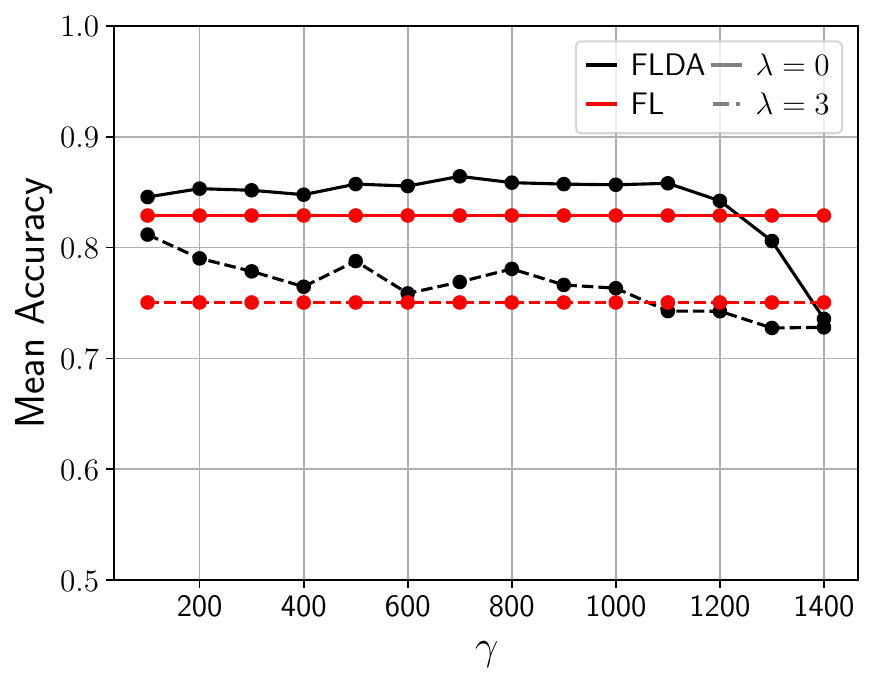} 
    \caption{Mean test accuracy as a function of $\gamma$ with the time fix at $150$ s for FL and the FLDA curves.}
    \label{fig:hybrid_accuracy_vs_gamma}
\end{figure}

In Fig.~\ref{fig:hybrid_accuracy_vs_gamma}, we fix the time at $150$ s and set $\alpha = 0.5$ to examine how the mean test accuracy of the device models in the FLDA method changes with $\gamma$. The FD curve is excluded from this analysis as it does not achieve accuracy levels comparable to FLDA under the same conditions. When the network operates without background traffic, the accuracy achieved by the device models is fairly constant, with a turning point at $\alpha\gamma = 500$. Beyond this turning point, the consensus phase becomes so long that the FL phase cannot converge to a global model, requiring a smaller $\alpha$, i.e., more  FL iterations compared to FD iterations within a full cycle. In contrast, the accuracy achieved by the models on the devices under high background traffic loads (e.g., $\lambda = 3$) does not remain constant.  In this case, the mean accuracy of the models on the devices decreases as $\gamma$ increases. This decline occurs because the devices spend more time performing FL iterations during each cycle, where their local updates are frequently lost due to collisions with the background traffic.
\captionsetup{skip=0pt}

\begin{figure}[t] 
    \centering
    \includegraphics[width=0.45\textwidth, trim=0mm 0mm 0mm 0mm,clip]{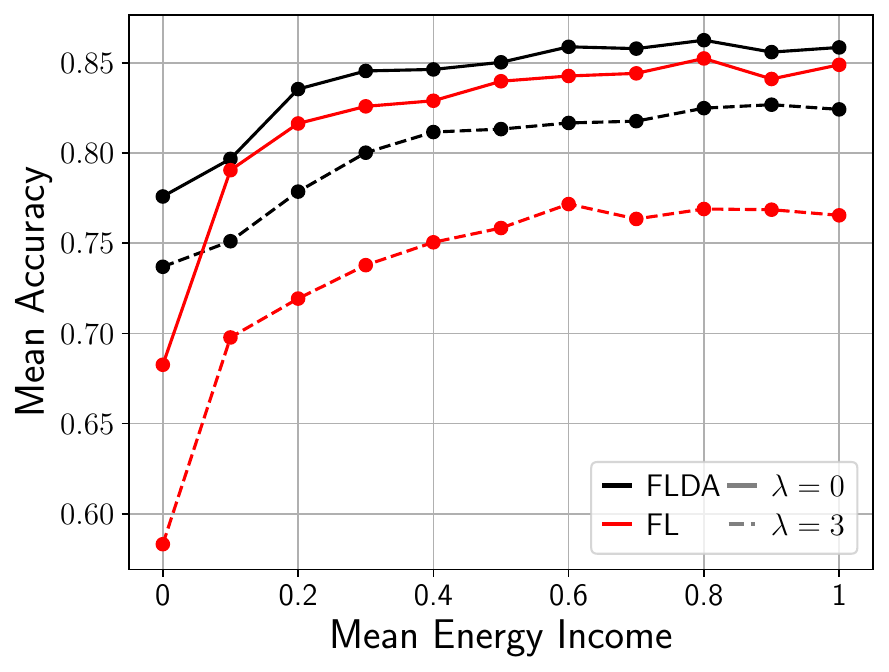} 
    \caption{Mean test accuracy as a function of time with the time fix at $150$ s for FL and the FLDA curves.}
    \label{fig:hybrid_accuracy_vs_meanEnergyIncome}
\end{figure}

We analyze the behavior of the mean accuracy as a function of the mean energy income in Fig.~\ref{fig:hybrid_accuracy_vs_meanEnergyIncome}. For this, we fix the time at $150$ s for FL and the proposed approach. We disregard FD in this analysis, as it fails to reach competitive accuracy levels compared to the proposed method. The results show that the models of both the proposed approach and FL achieve higher accuracy as the energy income increases. However, the faster convergence of the model in the proposed approach causes it to show a weaker dependence on~$\varrho$. This indicates that the proposed method is more robust in challenging EH-IoT scenarios, as its model performance remains less sensitive to energy income fluctuations.

\begin{table}[ht!]
\centering
\caption{Mean cumulative energy cost to reach different accuracy goals.}
\label{tab:CumulativeEnergy_for_AccuracyGoal}
\begin{tabular}{lcccc}
\toprule
\textbf{} & \textbf{Accuracy Goal} & \textbf{FLDA} & \textbf{FL}  & \textbf{Energy Savings} \\
\midrule
\multirow{3}{*}{$\lambda = 0$}   & $60$\% & \textbf{0.002 J} & 0.061 J  & 96.72\% \\
                                 & $70$\% & \textbf{0.016 J} & 0.096 J  & 83.33\%  \\
                                 & $80$\% & \textbf{0.089 J} & 0.181 J  & 50.82\% \\
\midrule
\multirow{3}{*}{$\lambda = 1.5$} & $60$\% & \textbf{0.002 J} & 0.066 J  & 96.97\%  \\
                                 & $70$\% & \textbf{0.016 J} & 0.102 J  & 84.31\% \\
                                 & $80$\% & \textbf{0.089 J} & 0.182 J  & 51.10\%  \\
\midrule
\multirow{3}{*}{$\lambda = 3$}   & $60$\% & \textbf{0.002 J} & 0.101 J  & 98.02\%  \\
                                 & $70$\% & \textbf{0.016 J} & 0.171 J  & 90.64\%  \\
                                 & $80$\% & \textbf{0.089 J} & - J  & - \%  \\\\
\end{tabular}
\end{table}
Finally, we examine the mean cumulative energy consumption required to reach specific accuracy targets for FL and FLDA, across three different background traffic loads and with \mbox{$\varrho = 0.4$}. We excluded FD from this analysis for two reasons: it does not achieve accuracies beyond $60$\%, and it cannot exceed the proposed method in energy efficiency when an proper $\alpha\gamma$ is chosen. In this case, we fix $\alpha= 0.5$ $\gamma$ and vary $\gamma$ from $100$ to $1400$ in steps of $100$. The results with higher energy savings are shown in Table~\ref{tab:CumulativeEnergy_for_AccuracyGoal} and evince that the proposed method consistently outperforms FL, demonstrating superior energy efficiency. In addition, energy savings are higher at lower accuracy targets, where the FD iterations are more relevant, and for increasing background traffic loads, where FL uplink throughput starts to experience significant degradation.

\section{Conclusion}
\label{sec:conclusion}
In this work, we studied the performance of FD and FL in resource-constrained EH-IoT scenarios considering multichannel SA under background traffic activity. The compromises between FD and FL under these conditions motivated the proposal of a new method, named FLDA, that repeatedly switches between FD and FL phases. FLDA was shown to outperform FD in terms of overall accuracy, and FL in terms of convergence time and overall accuracy. Moreover, with respect to FL, our proposal is less sensitive to high background traffic and uses less energy to reach target accuracies.

\section{Acknowledgment}
We thank Jihong Park for his insightful comments that helped shape this research, particularly with regard to the presentation of ideas in the paper.
\bibliographystyle{IEEEtran}
\mybibliography
\end{document}